# Vacancy-mediated complex phase selection in high entropy alloys

Prashant Singh[1], Shalabh Gupta[1], Srinivasa Thimmaiah[1], Bryce Thoeny[1,2], Pratik K. Ray[1,2], A.V. Smirnov,[1] Duane D. Johnson[1,2] and Matthew J. Kramer[1,2]

[1] Ames Laboratory, United States Department of Energy, Ames, IA 50011, USA

[2]Department of Materials Science & Engineering, Iowa State University, Ames, IA 50011, USA

**Abstract:**

Phase selection in Ti-Zr-Hf-Al high-entropy alloys was investigated by *in-situ* high-energy X-ray diffraction, single-crystal X-ray diffraction, and density-functional theory based electronic-structure methods that address disorder and vacancies, predicting formation enthalpy and chemical short-range order (SRO). Samples with varying Al content were synthesized, characterized, and computationally assessed to ascertain the composition-dependent phase selection, as increased Al content often acts as a stabilizer of a body-centered-cubic structure. Equiatomic TiZrHfAl was especially interesting due to its observed bcc superstructure – a variant of γ-brass with 4 vacancies per cell (not 2 as in γ-brass). We highlight how vacancy ordering mediates selection of this variant of γ-brass, which is driven by vacancy-atom SRO that dramatically suppress all atomic SRO. As vacancies are inherent in processing refractory systems, we expect that similar discoveries await in other high entropy alloys or in revisiting older experimental data.

**Introduction**

Entropic stabilization of disordered structures in multi-principal element alloys has instigated an entirely new materials arena, namely, High Entropy Alloys (HEAs).[1–7] Primarily the main focus has been on the role of configurational entropy, with relatively scant attention on the role of defects and chemical short-range order, which can be critical.[8] In addition to structure, chemical defects often affect a host of properties. For instance, anti-sites in NiAl[9–11] induce a significant width of the β-NiAl phase. Also, defects have a strong bearing on the mechanical behavior of off-stoichiometric intermetallics.[12] As for chemical disorder, HEAs containing refractory elements are expected to show creep behavior superior to Ni-based alloys or nickel-aluminides. Furthermore, our calculations in, e.g., TiZrHf indicate the presence of a body-centered-cubic (bcc) type structure at elevated temperature, similar to β-NiAl.[13] Hence, an assessment on the role of chemistry and defects, vacancies in particular, in these refractory-based materials is important, both from scientific and engineering perspectives. What structure will be adopted as a consequence of short-range order driven by addition of Al? How do vacancies, which naturally occur in processing in refractory-based systems, affect phase selection?

An intuitive understanding of the role of Al content on the formation of ordered structures is apparent if we look at the known binary and ternary aluminides in this system. The stronger bonding induced by Al is expected to yield an ordered structure, while a high configurational entropy for near-equiatomic compositions will tend to favor a solid-solution. Moreover, a bcc structure over hexagonal-closed-packed (hcp) should be more favored entropically at higher temperatures due to higher vibrational entropy. The aluminides in Al-Zr and Al-Hf systems form line compounds, while those in Al-Ti have a significant phase width due to the presence of anti-site defects.[14] Equiatomic AlHf congruently melts, crystallizing with a TlI-type structure (*Cmcm* space group),[15] while AlZr has the same structure as ZrHf but dissociates[16] above 1275°C to $Zr_2Al_3$ and $Zr_5Al_4$. In contrast, AlTi crystallizes in a face-centered-cubic (fcc) CuAu ($L1_0$) structure over a range of composition.[17] Relatively few reports exist for equiatomic ternaries containing Al. In TiHfAl, a three-phase microstructure is found (hcp-TiHf with HfAl+TiAl), although lower Al content (<12-25 *at.%Al* in atomic percent, depending on *Ti* or *Hf* concentrations ) exhibits a single-phase hcp solid solution. TiZrAl shows the presence of a $Ti_{50}Zr_{25}Al_{25}$ B2-phase, with lower *Al* concentration a hcp solid solution is present.

Here, we illustrate the role of vacancies in structure selection with a systematic study using diffraction experiments and density-functional-theory (DFT) phase-stability analysis,[18] calculating both formation enthalpies and short-range order (chemical- and vacancy-ordering modes) in A1 (fcc), A2 (bcc), and A3 (hcp) solid-solution phases. The TiZrHfAl$_x$ ($0 \leq x \leq 1$ atomic-fraction; or 0–25 *at.%Al*) is adopted as a model system to identify factors dictating changes of stability. The addition of Al is anticipated to induce a structural change beyond a critical content, as Al often plays the role of bcc stabilizer, as found in Cantor alloys.[8] A sweep through composition space indicates relatively small number of alloys with better thermodynamic stability on increasing *Al* concentration. From theory, we address the potential for vacancy-stabilized structures through the vacancy formation energy in bcc $\square_y$(TiZrHfAl)$_{1-y}$, where *y* is the fraction of vacancies (denoted by $\square$). That is, vacancies are treated as another "alloying element" that may exhibit disorder, short-range order (SRO), or long-range order. The stabilization due to vacancy ordering is established through single-crystal diffraction and is further corroborated by Rietveld analyzed synchrotron data. And, the physics of vacancy-mediated complex structure stabilization is explained by comparing the theoretical energetics and chemical SRO (showing the incipient vacancy ordering) that results from electronic structure of a disordered phase containing vacancies, as also compared to the observed vacancy-stabilized structure. We find that vacancies form an array that stabilizes a complex structure in HEAs – a variant of γ-brass having 4 vacancies per unit cell (not 2, as in traditional brasses) – rather than an expected simpler cubic or hexagonal solid-solution.

**Methods**:

*Materials Preparation and Characterization*: Samples were prepared by arc melting and casting on a water-cooled copper hearth in Ar from elemental metals (Alfa Aesar, purity > 99.8%). Five different alloys with the composition TiZrHfAl$_x$ (x = 0, ¼, ½, ¾, 1) were prepared. Samples were melted and flipped three times to insure homogeneity. The XRD data was obtained at the Advanced Photon Source. The room temperature data on ground powder (contained in a Kapton capillary) was obtained at sector 11B with 88.0 keV photons using a 2D solid-state detector at 30 and 85 mm from the sample. The data taken at 30 mm was used to obtain pair distribution function data and the 85 mm was used for standard Rietveld analysis. High-temperature data was obtained at sector 11C with 105.6 keV. Ground powders could not be used due to the measurement geometry. Hence, thin slices (< 500μm) of the alloy, sectioned from the arc-melted buttons, were used for high-temperature studies. Two to three slices were used for each high temperature run to ensure improved powder averaging, i.e., the samples were cut into thin slices and stacked two to three on top of each other. Heating was done with a Linkham Pt wound resistance heater in a flowing Ar-environment. $CeO_2$ was used to calibrate the diffraction data. The synchrotron data was analyzed using the GSAS-II package.[19] Additionally, for the five compositions, scanning electron microscopy (SEM) at three magnifications (lowest being 5000 ×) and energy-dispersive spectroscopy (EDS) were performed on as-cast samples. The alloy densities were determined using X-ray data of as-cast $TiZrHfAl_x$ samples.

*Single crystal data*: Single crystal diffraction studies were done using several irregular shaped crystals from a nominal composition of TiZrHfAl alloys, which were tested and measured on a STOE IPDS II single crystal diffractometer. X-ray diffraction intensities were collected with the use of graphite monochromatized Mo-Kα radiation (λ = 0.71073 Å) at room temperature. The data reduction and processing were carried out using Stoe X-Area program package. The measured intensities were corrected for Lorentz and polarization effects, and a numerical absorption correction was accomplished with the program X-SHAPE.

The unit cell parameters and Bravais lattice were obtained by indexing peaks from 3840 measured reflections. A careful examination of the systematic absence reflections indicated an acentric, primitive cubic system with the Laue group $m\bar{3}m$. Finally, the space group $P\bar{4}3m$ was chosen for subsequent structural solution and refinement. The structural model was obtained from direct methods. Subsequent structural refinement was carried out on $F_2$ using full-matrix least-squares procedures using the SHELXTL package. After careful examination of several different structure models, we choose the one which consists of nine crystallographic independent positions. Since this model provided meaningful interatomic distances and lower residual R1 value (R1 = 26%) compared to others. Accurate assignment of atoms based on thermal parameters could not be accomplished due to strong correlation between the displacement parameters and high degree of mixed occupancies between different pair of atoms. A slight improvement in R1 value (R1 = 21%) was achieved by refining the structure with mixed occupancies. The structure model obtained from the single crystal refinement serves as a starting model for X-ray powder refinement as well as electronic structure calculations.

***Density-Functional Theory Method***: Formation enthalpies of TiZrHfAl$_x$ (x ≤ 1 atomic-fraction, or *at.%Al* ≤ 25) in competing phases were calculated using DFT-based electronic-structure methods,[20–22] in which the coherent-potential approximation (CPA) accounts properly for chemical disorder in structures with/without defects, such as vacancies or stacking faults.[23–25] We employed a gradient-corrected exchange-correlation functional (PBESol),[26] and use an atomic sphere approximation for each site's Voronoi polyhedral. Yet, we account for periodic boundary conditions and spatial integrals over Voronoi polyhedra for unit cell charge distribution and energy calculation. To calculate the Green's functions, a semi-circular (complex-energy) contour integration used a Gauss-Laguerre quadrature with 24 complex energies (enclosing the bottom to the top of the valence states) and Brillouin-zone integrations used a (24 × 24 × 24) **k**-space mesh (Monkhort-Pack type). For each site, the KKR basis employed a $L_{max} = (\ell, m) = 3$ spherical harmonic basis which includes *s, p, d*, and *f*-electron symmetries, and an automated choice of ASA radii, with details in Refs. 22-25. The self-consistent charge and energy tolerances for convergence were set at 10$_{-7}$ eV/atom and 10$_{-6}$ eV/atom, respectively, for total alloy energy and ground-state volume calculations. A Birch-Murnaghan equation of state was used to fit energy vs volume (or lattice-constant). The *c/a* of 1.58564 is used for hcp phase from experiment.

*Vacancy calculation*: To understand the relative stability of this γ-brass-type phase, we performed DFT calculations using a 3 × 3 × 3 supercell (54-atoms) with observed lattice constant (10.196 Å) with 4 "ideally-located" atoms removed (vacancies), with no relaxations to facilitate identification of key electronic features. We note that γ-brass-type phase can be defined within a 3 × 3 × 3 supercell of the cubic bcc (2-atom) cell (see more discussion in Section I of the supplement).

*Formation energy calculation*: The formation energy (*E*$_{form}$) of Ti-Zr-Hf-Al system was calculated using $E_{form} = E_{total}^{alloy}(\{c_i\}) - \sum_{i=1,N} c_i E_i$, where $E_{total}^{alloy}$ is the total energy of the alloy (with or without vacancy), $c_i$ is the composition of alloying elements, $E_i$ is the elemental energy, and '*i*' labels elements (Ti, Zr, Hf, and Al). For *E*$_{form}$, hcp and fcc are reference phases of (Ti/Zr/Hf), and Al, respectively.

***Short-Range Order (SRO) Theory***: We calculated Warren-Cowley SRO parameters $\alpha_{\mu\nu}(\mathbf{k}; T)$ using a unique DFT-based KKR-CPA thermodynamic linear-response method.[8,27] The SRO pair correlations (chemical modes) that develop in solid-solutions arise from correlated fluctuations in site-occupation probabilities (i.e., site concentrations) and can be calculated using linear-response, as done for phonon (or site-displacement) modes to get the "force-constant" (vibrational stability) matrix. The Warren-Cowley SRO in an alloy is dictated by $\mu-\nu$ pairwise interchange energies $[S^{(2)}_{\mu\nu}(\mathbf{k};T)]$ as prescribed by the electronic structure of the alloy, i.e.,

$$[\alpha^{-1}(\mathbf{k};T)]_{\mu\nu} = \mathbb{C}_{\mu\nu} - (k_B T)^{-1} c_\mu (\delta_{\mu\nu} - c_\nu) S^{(2)}_{\mu\nu}(\mathbf{k};T) \quad , \tag{1}$$

with $k_B$ Boltzmann's constant and T temperature in Kelvin.[28] Here, $\mathbb{C}_{\mu\nu} = c_\mu(\delta_{\mu\nu} - c_\nu)(\delta_{\mu\nu}/c_\mu + 1/c_N)$ is a matrix element involving N compositions $c_\nu$, with N[th] element taken as reference (or "host").[8,28,29] The thermodynamically-averaged quantity $S^{(2)}_{\mu\nu}(\mathbf{k};T)$ is the chemical stability matrix of the alloy; this interchange energy is a $(N-1) \times (N-1)$ matrix given in linear-response theory by a second-variation of free-energy using the average (scattering) lattice of the homogeneous solid-solution as reference state,[20] as experimentally defined. The diffuse scattering intensities are given in terms of Laue units,[8,27] a Laue unit is defined as $c_\mu(\delta_{\mu\nu} - c_\nu)(f_\mu - f_\nu)^2$, where $f_\mu, f_\nu$ are the atomic scattering form factors, and $(\mu,\nu)$ are component labels, and $(c_\mu, c_\nu)$ are the compositions. Hence, the intensity is $I(\mathbf{k};T) = \sum_{\mu,\nu} c_\mu(\delta_{\mu\nu} - c_\nu)(f_\mu - f_\nu)^2 \alpha_{\mu\nu}(\mathbf{k};T)$. And, by this definition, the Warren-Cowley SRO parameters are properly defined as normalized pair-probabilities. A discussion of relaxations and why the average lattice is appropriate to calculation of SRO in linear-response theory is provided in the Supplement.

Now, regarding Eq. (1), the most unstable SRO mode has the largest peak in $\alpha_{\mu\nu}(\mathbf{k}_o; T > T_{sp})$ in the occurring at wavevector $\mathbf{k}_o$ for a specific $\mu$-$\nu$ pair in the solid-solution phase. An absolute instability to $\mathbf{k}_o$ modes[8,27] occurs below the spinodal-decomposition temperature $T_{sp}$, where $[\alpha^{-1}(\mathbf{k}_o; T_{sp})]_{\mu\nu} = 0$. For $\mathbf{k}_o = (000)$, the alloy is unstable to segregation (atomic or vacancy clustering), rather than local ordering. And, it is possible that both finite $\mathbf{k}_o$ and $(000)$ peaks compete. Notably, if $\mathbf{k}_o$ peaks are not at high-symmetry points in the Brillouin zone (indicating small unit cell ordering), it will peak along high-symmetry lines indicateing Fermi-surface nesting effects are operative.[28]

**Results and Discussion**

*Phase Stability*: In **Figure 1**, the DFT-calculated formation enthalpies ($E_{form}$) provide relative energies of A1 (fcc), A2 (bcc), and A3 (hcp) phases. The high configurational entropy in an N-component equiatomic HEA (i.e., $\Delta S \sim -k_B \log N$) added to $E_{form}$ to form the free energy aids in formation of solid-solution phases. Thermodynamic assessment of TiZrHf (x=0) system[13] indicates the presence of an A3 phase at room temperature, which transforms to A2 above 850°C. So, at elevated temperatures near x~0, an energetically similar but more open bcc structure with higher vibrational entropy is preferred (as observed) over the closed-packed solid solution. The constituent binaries (TiZr, TiHf, and ZrHf) show complete solubility in α (A3) and β (A2) allotropic forms; hence, it is no surprise that the room-temperature ternary exists as α phase. Clearly, $E_{form}$ (**Fig. 1**) indicates that increasing *Al* concentration (*x*) for TiZrHfAl$_x$ stabilizes the A2 phase, i.e., *Al* plays the role of bcc stabilizer, as found in Cantor alloys, where *Al s-p* band better hybridizes with bcc transition-metal *d*-states.[8] The two phase region in Fig. 1 arrives when chemical potentials of two phase become equal.[30] We also perform supercell calculation for equiatomic TiZrHfAl using 128 atom ($4 \times 4 \times 4\ a_{bcc}$; Ti=32; Zr=32; Hf=32; Al=32) special-quasi random structure supercell[31] within VASP[32] with PBESol exchange-correlation potential.[26] The calculated $E_{form}$ of fully relaxed supercell is -134.37 meV/atom, which agrees well with the KKR-CPA $E_{form}$ of -139.9 meV/atom in **Fig. 1**. KKR-CPA and VASP calculated lattice constant ($a_{bcc}$) per 2 atom per cell are 3.401 Å and 3.404 Å, respectively.

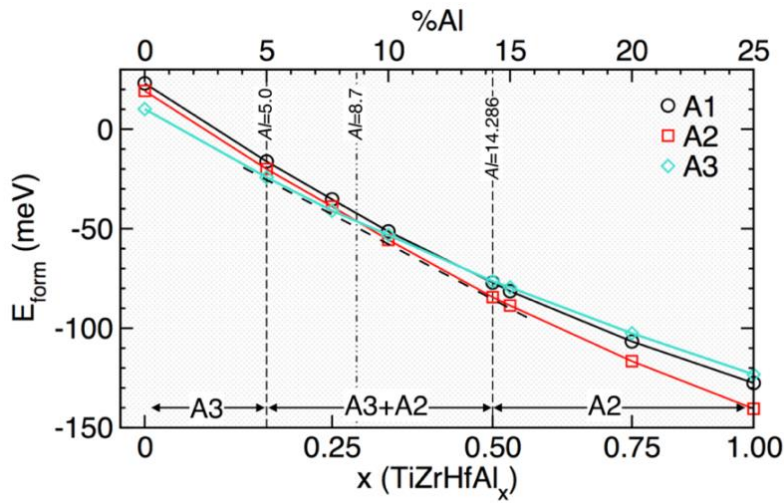

**Figure 1**. For A1/A2/A3 (TiZrHf)Al$_x$, formation energy (meV/atom) at 0 K versus x (atomic-fraction) or *at.%Al*. Increased *Al* concentration stabilizes A2, indicating a single A2 phase at higher *Al* concentration. Maxwell construction is used to identify single-phase and two-phase regions, as indicated. Note that y=x/(3+x), where x is the atomic fraction of Al.

To verify the phase stability calculations, we synthesized samples of TiZrHfAl$_x$ with varying Al content (x = 0, ¼, ½, ¾, 1) by arc-melting. The synchrotron XRD patterns collected from these as-cast samples are shown in **Fig. 2a**. As *Al* concentration is increased, we observed a gradual shift in peaks to smaller unit cell size for $x \leq 0.5$ atomic-fraction in **Table 1**. All compounds for x ≤ 0.5 crystallized with a major A3 phase ($P6_3/mmc$, as seen in refractories and their binaries). The powder patterns showed increased peak broadening with higher *Al* concentration. For atomic-fraction x ≥ 0.75 (i.e., 20 and 25 *at.%Al*), a new phase formed in a bcc superstructure with a γ-brass variant ($P\bar{4}3m$), as found by synchrotron powder XRD and single-crystal diffraction for x=1.

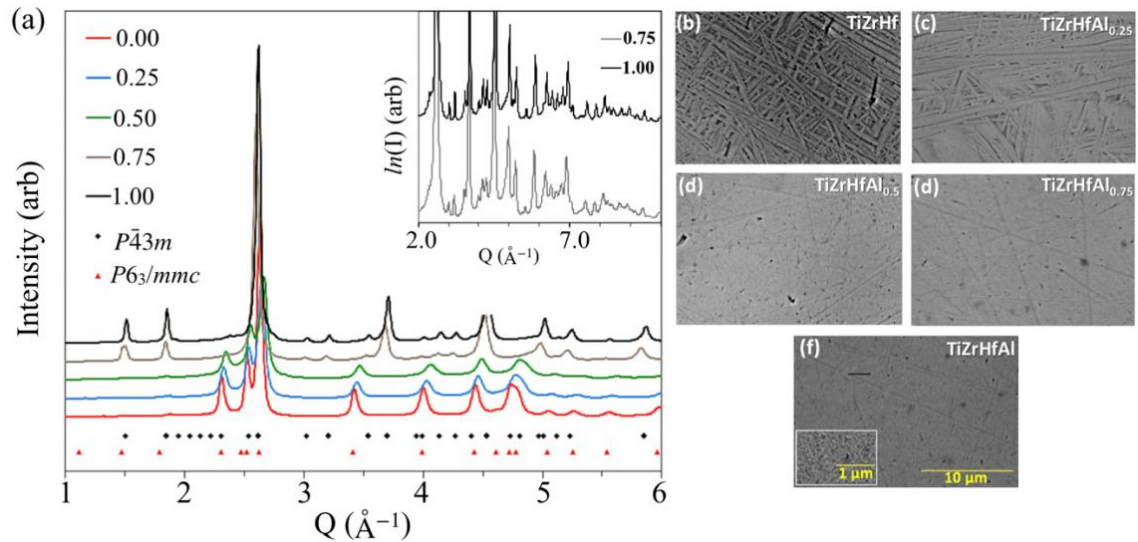

**Figure 2.** (a) X-ray diffraction intensities in as-cast arc-melted TiZrHfAl$_x$ vs. *x* (atomic-fraction), showing its effect on phase selection. At *x* < 0.75, a single-phase A3 is observed at 300 K, where the main A3 peaks vary with increasing *Al* (expected by diluting solid-solution). Increasing *Al* destabilizes A3 and a new phase is formed, identified as a variant of γ-brass structure (a bcc superstructure). See Table 1. **Inset:** Intensity (log-scale) shows a several minor peaks consistent with ordered phase for x=0.75 and 1. (b-f) SEM micrographs (backscattered electron images) confirms that all five samples are in single solid-solution phase (with measured compositions very close to the targeted composition, see **Table 2**, and **Fig. S1-S5**).

Our SEM and EDS analyses (**Fig. 2b-2f**, **Table 2, and Fig. S1-S5**) confirmed the samples to be nearly in single solid-solution phase, with the measured compositions very close to the targeted composition (within limitations of the technique). The samples did show increasingly finer grain size with increasing Al and the backscattered electron images show a higher Z content around the grain boundaries for samples for x ≥ 0.75 (inset **Fig. 2f** and **Fig. S4-S5**). The phase stability of as-cast alloys was also investigated by *in situ* high-energy powder diffraction to 1100°C on heating and cooling at 20 °C/min (**Fig. S6**). This provides further insight into the solid-solution and stability, especially atomic and vacancy-atom SRO, possibly prevalent in processing of refractory metals. For $x \leq 0.5$, the samples partially converted (~20 *at.%*) to the A2 ($Im\bar{3}m$) above ~950°C, which reverted to A3 upon cooling, although at a much lower temperature. Minor secondary phases were more prevalent with increasing *Al* concentration (small volume fractions of high-Z at the grain boundaries), especially after annealing.

**Table 1**. Lattice parameters (Å), volume (Å³), space group and density for as-cast samples from synchrotron XRD powder diffraction and fitting parameters. For A3, the composition of the single 2c site $(\frac{1}{3}, \frac{2}{3}, \frac{1}{4})$ was fixed to that in bulk. See text for details for atomic positions in $\gamma$-brass ($P\bar{4}3m$) in Table 3. The density is based on the X-ray volume.

|      | a-axis (Å) | c-axis (Å) | Volume (Å³) | Space group | Density (g/cc) | RF₂ |
|------|------------|------------|-------------|-------------|----------------|-----|
| **0.00** | 3.148(6)  | 4.986(3)  | 42.81(1)    | $P6_3/mmc$  | 8.21(3)        | 2.26%, 5.77% on 52 reflections |
| **0.25** | 3.130(7)  | 4.963(8)  | 42.13(3)    | $P6_3/mmc$  | 7.86(7)        | 2.68%, 7.89% on 55 reflections |
| **0.50** | 3.110(8)  | 4.937(1)  | 41.37(6)    | $P6_3/mmc$  | 7.59(3)        | 4.07%, 13.92% on 51 reflections |
| **0.75** | 10.272(3) |            | 1083.94(9)  | $P\bar{4}3m$ | 6.90(1)       | 8.06%, 6.67% on 594 reflections |
| **1.00** | 10.195(8) |            | 1059.97(1)  | $P\bar{4}3m$ | 6.74(8)       | 8.61%, 6.72% on 574 reflections |

**Table 2.** For TiZrHfAl$_x$, measured average compositions over a mapped area (~10 × 10 μm) compared to nominal (normalized) compositions, showing a variation within ~1 $at.$% of a point analysis (4 to 5 points).

| Nominal | Nominal (normalized, at. %) | Measured (averaged, at. %) |
|---------|-----------------------------|----------------------------|
| TiZrHfAl$_{0.00}$ | Ti$_{0.33}$Zr$_{0.33}$Hf$_{0.33}$ | Ti$_{32(1)}$Zr$_{33.5(5)}$Hf$_{34(1)}$ |
| TiZrHfAl$_{0.25}$ | Ti$_{30.77}$Zr$_{30.77}$Hf$_{30.77}$Al$_{7.69}$ | Ti$_{32(2)}$Zr$_{29.6(7)}$Hf$_{30(2)}$Al$_{8.2(2)}$ |
| TiZrHfAl$_{0.50}$ | Ti$_{28.57}$Zr$_{28.57}$Hf$_{28.57}$Al$_{14.28}$ | Ti$_{27.3(2)}$Zr$_{28.34(8)}$Hf$_{30.2(2)}$Al$_{14.1(1)}$ |
| TiZrHfAl$_{0.75}$ | Ti$_{26.66}$Zr$_{26.66}$Hf$_{26.66}$Al$_{20.0}$ | Ti$_{26.2(6)}$Zr$_{27.2(3)}$Hf$_{27.8(2)}$Al$_{18.8(4)}$ |
| TiZrHfAl$_{1.00}$ | Ti$_{25.0}$Zr$_{25.0}$Hf$_{25.0}$Al$_{25.0}$ | Ti$_{25(1)}$Zr$_{26.4(8)}$Hf$_{26.2(4)}$Al$_{22(1)}$ |

*Structural Stability and SRO*: To identify mechanism(s) responsible for formation of a γ-brass variant over bcc TiZrHfAl with 25 *at.%Al* (**Fig. 1**), we investigated the role of Al concentration and vacancy concentration on the atomic SRO. Notably, we include vacancies (denoted by ▫) as another "element", i.e., ▫$_y$(TiZrHfAl)$_{1-y}$, giving 6 *atom-atom pairs* plus 4 *vacancy-atom* pairs that can drive SRO and vacancy-induced ordering.

In **Figure 3**, for equiatomic bcc TiZrHfAl, we present calculated Warren-Cowley SRO parameters, i.e., $\alpha_{\mu\nu}(\boldsymbol{k};T)$ in Laue units. Warren-Cowley pair correlations are measured relative to the average X-ray lattice, as atomic displacements sum to zero on average (by symmetry, for each spatial direction), so we evaluate the SRO at the calculated thermodynamic average lattice of the homogeneous alloy. While atomic displacements exhibit a distribution about zero, they do not affect the average volume (by symmetry), the average electronic properties, nor the symmetry of chemical SRO (as shown long ago experimentally by Borie,[33] albeit the resulting Debye-Waller factors do broaden intensities. For bcc TiZrHfAl with no vacancies, the calculated SRO in **Fig. 3(a)** exhibits a peak at $\boldsymbol{k_o}=\boldsymbol{H}$=<111>, giving B2 (CsCl-type) SRO with an instability to partial long-range order below $T_{sp}$ of 730 K. The Hf-Al pairs in **Fig. 3(a)** are the principal driver for SRO, as dictated by $S^{(2)}_{\mu\nu}(\boldsymbol{k}_o = (111);T)$ in **Fig. 3(a')**. A mean-field estimate from our calculated B2-A2 energy difference, i.e., $k_B T_{o-d} \sim E_{form}(B2) - E_{form}(A2) = -65.3\ meV$, gives an order-disorder transition at 757 K, thermodynamically consistent to $T_{sp}$ of 730 K from linear-response.

Remarkably, with vacancies, i.e., ▫$_y$(TiZrHfAl)$_{1-y}$, the atomic SRO is suppressed, **Fig. 3(a-d)**, where peaks at **H** are reduced rapidly with vacancy concentration: 15 Laue (0 *at.%Vac*); 8 Laue (2 *at.%Vac*); 2 Laue (4 *at.%Vac*) to 1.5 Laue (8 *at.%Vac*). Although weakening with vacancy concentration, the aluminum-refractory pairs Zr-Al, Ti-Al, and Hf-Al SRO remains B2-type (peak at **H**), i.e., atomic pairs want to be on different neighboring sites, while refractory-metal pairs Ti-Zr, Ti-Hf, and Zr-Hf remain, more or less, mixed, which affect the preferred structure. By 2 *at.%Vac*, the atomic SRO and Vac-Ti SRO peak at **H** are comparable with B2-type *atomic* and *vacancy-atom* SRO, driven by $\boldsymbol{S}^{(2)}_{Vac-Zr}(\boldsymbol{\Gamma})$, while Vac-Zr and Vac-Hf SRO exhibit vacancy clustering (segregation). Then, above 2 *at.%Vac*, *vacancy-atom* SRO in **Fig. 3(b$_1$–d$_1$)** dominates over all *atomic* SRO in **Fig. 3(b-d)**. At 4 *at.%Vac* and above, *vacancy-atom* SRO now exhibits segregation for Ti, Zr, and Hf that intensifies.

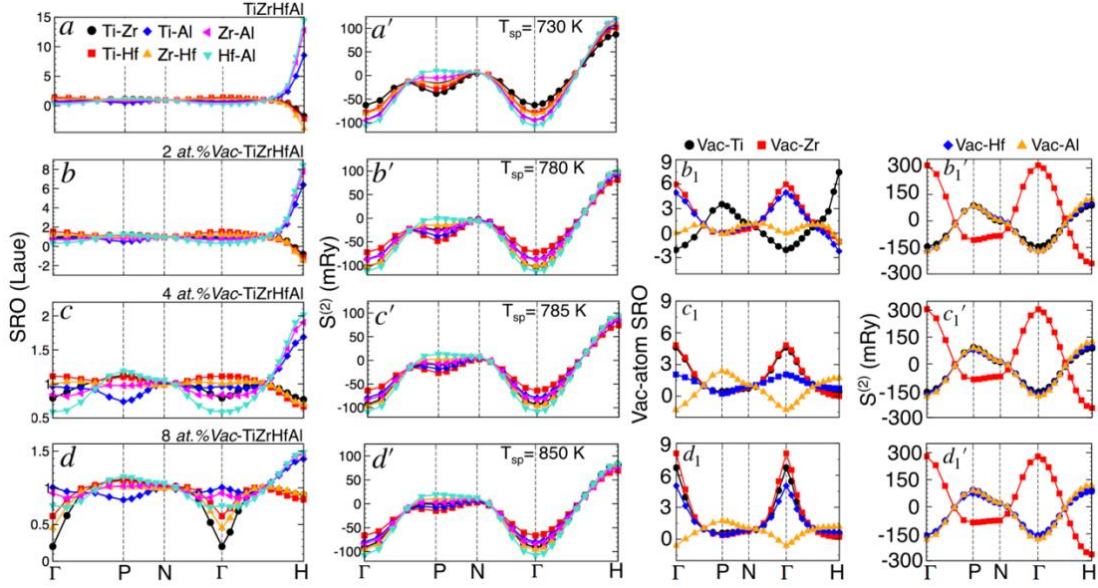

**Figure 3**. SRO for $\square_y$(TiZrHfAl)$_{1-y}$ plotted along Brillouin zone high-symmetry directions. (a) $\alpha_{\alpha\beta}$(k; T=1.15T$_{sp}$), and (a') pair-interchange energies, $S^{(2)}_{\alpha\beta}$(k), for all 6 independent pairs. SRO peaks at **H**=[111], reflected in $\alpha_{\alpha\beta}$(**H**) for aluminum-refractory pairs (Hf-Al, Zr-Al, and Ti-Al) but driven by specific $S^{(2)}_{\alpha\beta}$(**H**). For y = 2, 4, and 8 *at.%Vac*, (b–d) $\alpha_{\alpha\beta}$(k; T=1.15 T$_{sp}$) and (b'–d') $S^{(2)}_{\alpha\beta}$(k) with T$_{sp}$ of (b, b') 780 K; (c, c') 785 K; (d, d') 850 K, indicating larger instability driven by *Vac-Atom* SRO (right-hand side), manifest as Vac-Ti order at 2 *at.%Vac* and then segregate from Vac-Zr pairs for 4 *at.%* to 8 *at.%Vac*.

As expected, this vacancy-segregation correlates directly with the solid-solution $\square_y$(TiZrHfAl$_x$)$_{1-y}$ $E_{form}$ versus vacancy concentration (**Fig. 4**), which is linear and positive above ~3.7 *at.%Vac*, as vacancy defects always cost energy. Hence, the solid-solution goes from $E_{form}$ < 0 (stable) to $E_{form}$ > 0 (unstable) with vacancies – so inherently a bulk alloy with greater than 3.7 *at.%Vac* (i.e., 2 vacant sites in 54-atom bcc cube) is energetically less stable with respect to an alternate structure with ordering of vacancies or vacancy clusters. Notice that the formation energies of γ-brass (evaluated at the experimental volume) equals that of the solid-solution at 6 at.%Vac (**Fig. 4**), nearing 4 vacancies per 54-atom cube. As noted above, the SRO (**Fig. 3**) indicates what incipient order is preferred in the solid-solution: aluminum-refractory pairs (Zr-Al, Ti-Al, and Hf-Al) want B2-type neighboring sites to satisfy SRO, while refractory-metal pairs (Ti-Zr, Ti-Hf, and Zr-Hf) remain mostly mixed. Indeed, diffraction analysis (below) giving assessed structure (**Fig. 4c**) reflects these pair correlations with prominence of B2-type vacancy ordering.

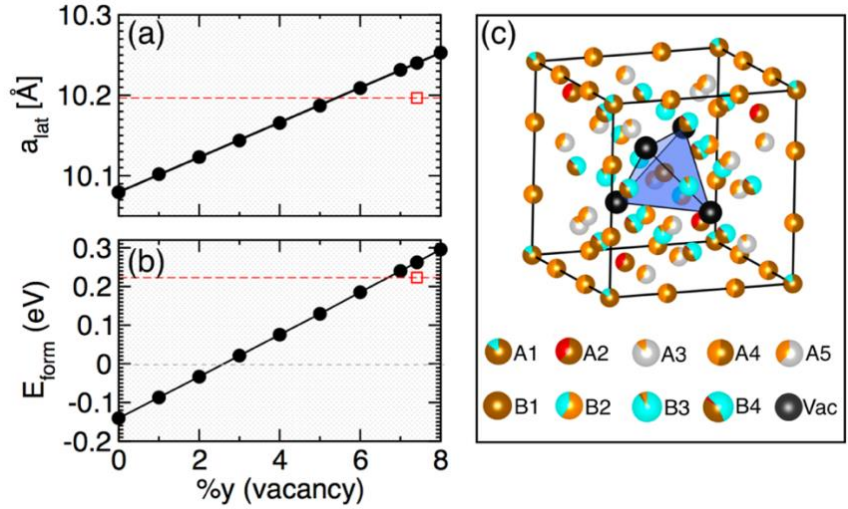

**Figure 4**. (a) Lattice constant (Å), and (b) formation energy (*eV/atom*) for homogeneously-distributed vacancies (0-8 *at.%Vac*) in bcc ☐$_y$(TiZrHfAl)$_{1-y}$. Results are compared to γ-brass structure (c) with vacancy tetrahedron (black) sites. See **Table S1** for computed site-occupancies. Measured γ-brass lattice-constant 10.196 Å (single-crystal values marked with red-square, dashed lines are guide to eyes) and calculated $E_{form}$ for γ-brass structure (c) compared to homogeneous HEA, e.g., 7.41 *at.%Vac* (4 per 54-atom cube, with 10.24 Å for powder-sample with lattice constant as 3$a_{bcc}$). The variant of γ-brass is described in a **3 × 3 × 3** supercell of the cubic bcc (2-atom) cell.

*Structure Determination*: From single-crystal diffraction data analysis, we derive a structure model of TiZrHfAl high entropy alloy and extract, e.g., lattice parameters and Bravais lattice. However, given the complexity of this structure and number of elements, the atomic positions could not be assigned accurately to any specific elements in the structure, initially expected to be a bcc solid-solution. The reflections are indexed in a primitive cubic lattice (space group $P\bar{4}3m$) with lattice constant *a* of 10.214 Å for the single-crystal and 10.196 Å for powder results (see **Table 1**). From refinement using powder data, TiZrHaAl has 50 atoms in a primitive cubic cell, distributed among nine crystallographically independent positions (**Fig. 5**). The decoration of atoms in the structure closely resembles many known γ-brass-type structures. The novel structure can be described using cluster or nested polyhedra to identify the topological similarity between the related structures. From positional parameters found in diffraction, atoms are arranged into 9 Wyckoff positions (**Table 3**). These 9 sites form two distinct clusters (**Fig. 5a**) centered on cube corners (27-atom cluster **A**) and centers (23-atom cluster **B**), respectively.

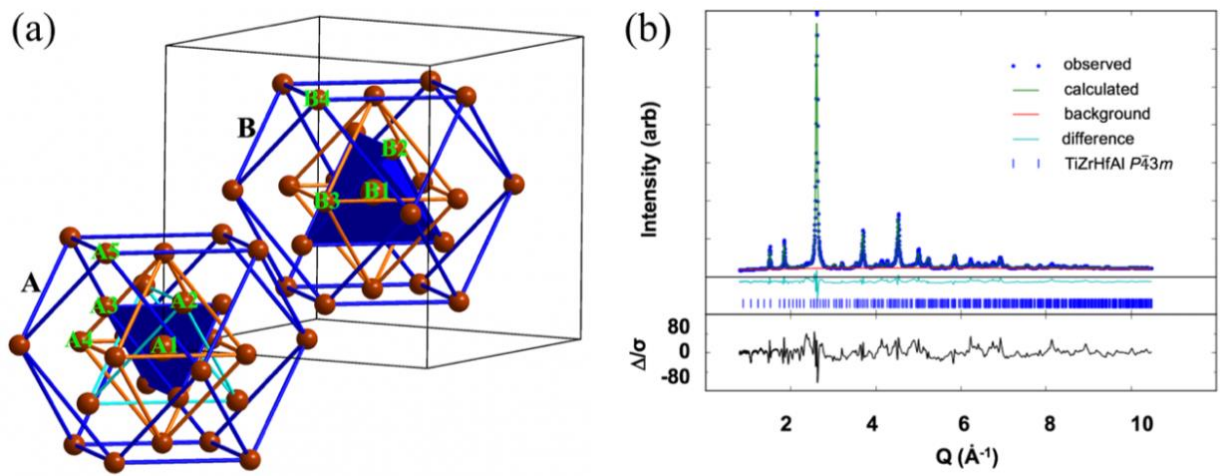

**Figure 5.** (a) Structure of TiZrHfAl as determined from single-crystal diffraction. The 50-atom unit cell is generated by 5 [A1, A2, A3, A4, A5] and 4 [B1, B2, B3, B4] inequivalent Wycoff sites (**Table 3**), reflecting two polyhedrons with 27 and 23 atoms, respectively, centered around A1=(0 0 0) and B1=(½ ½ ½) of the cubic unit cell; (b) Rietveld fitting of synchrotron data used structural model described in **Fig. 5a** and assessed in **Fig. 4c**.

In general, the γ-brass structure (space group $Im\bar{3}m$, or Pearson $cI$52) can be described as a bcc packing of 26-atom clusters, constructed by concentric shells of atoms centered about a vacancy at the 1$a$ Wyckoff site (0,0,0). The 26-atom clusters consist of an (a) inner tetrahedron; (b) outer tetrahedron; (c) octahedron; and a (d) distorted cuboctahedron. This description of γ-brass is robust and useful for direct comparisons with other known γ-brass structures in different systems to address site preferences for different elements in these phases. In several complex systems, the atoms in the clusters vary between 22 and 29 atoms, such as Ti$_2$Ni-type (22), α-Mn-type (29), bcc-type (27), or γ-brass (26).

For TiZrHfAl, cluster A is fully occupied by 27 atoms, but cluster B has an absence of inner tetrahedron (**4 vacancies**) around the central atom, giving only 23 atoms (**Fig. 5a** and **Table 3**). Distances between central and peripheral atoms (**Table 3**) is 4.98 and 4.87 Å for cluster A and B, respectively. Strong correlation between the stability of specific crystal structure and the valence electron count (VEC, or electron-per-atom ($e/a$) ratio) is frequently observed in Hume-Rothery compounds.[34] Among several phases, the complex γ-brass emerges[34,35] for VEC in a range of 1.59 to 1.75. The vacancies provide additional structural stability by reducing VEC in γ-brass phases.[36–38] The VEC-dependent *cubic*-to-*rhombohedral* distortion is well demonstrated for γ-Cu$_9$Al$_4$,[36] which crystallizes in a 52-atom P-cell. Increasing *Al* concentration in γ-Cu$_9$Al$_4$ creates additional vacancies and minimizes the VEC.36 A similar VEC effect was observed for superstructures of γ–brass (γ'-cF400-cF416) phases by forming vacancies or mixed atomic site occupancies to keep the VEC in a range of 1.59–1.75. This VEC analysis supports the existence of a new complex γ–brass phase in TiZrHfAl, formed by the presence of *4 vacant sites*, in contrast to *2 vacant sites* in traditional brass.

**Table 3.** For TiZrHfAl, high-symmetry positions from single-crystal diffraction. Cluster B is missing *4e* sites, i.e., location of the 4 vacancies (▫). The radius (R) is the distance from the cluster center to each site, with vacant positions estimated based on symmetry. Notably, A and B clusters are inequivalent based on both occupations and positions. Calculated elemental site occupancies of newly discovered $\gamma$-brass ($P\bar{4}3m$) phase are given in **Table S1**.

| 27-atom cluster A centered on A1 | | | | | 23-atom cluster B centered on B1 | | | | | |
|---|---|---|---|---|---|---|---|---|---|---|
| Site | Wyckoff | $x$ | $y$ | $z$ | $R$ (Å) | Site | Wyckoff | $x$ | $y$ | $z$ | $R$ (Å) |
| A1 | 1a | 0 | 0 | 0 | 0 | B1 | 1b | 0.5 | 0.5 | 0.5 | 0 |
| A2 | 4e | 0.1410 | 0.1410 | 0.1410 | 2.4946 | B2 | 4e | 0.3305 | 0.3305 | 0.3305 | 2.9988 |
| A3 | 4e | 0.8214 | 0.8214 | 08214 | 3.1598 | ▫ | 4e | 0.68 | 0.68 | 0.68 | 3.1846 |
| A4 | 6f | 0.3140 | 0.0000 | 0.0000 | 3.2074 | B3 | 6g | 0.1810 | 0.5000 | 0.5000 | 3.2585 |
| A5 | 12i | 0.3445 | 0.3445 | 0.0198 | 4.9806 | B4 | 12i | 0.1631 | 0.1631 | 0.5166 | 4.8697 |

To further support our analysis and new structure, we measured the density of samples versus Al content using X-ray, which then was compared to the theoretical densities with and without the vacancy tetrahedron (**Fig. 6**). As shown in the figure, if the theory includes the ordered vacancy-tetrahedron array, the densities match the abrupt drop in density upon entering the bcc phase in new complex γ–brass phase; if not, then the density should just show linear dependence.

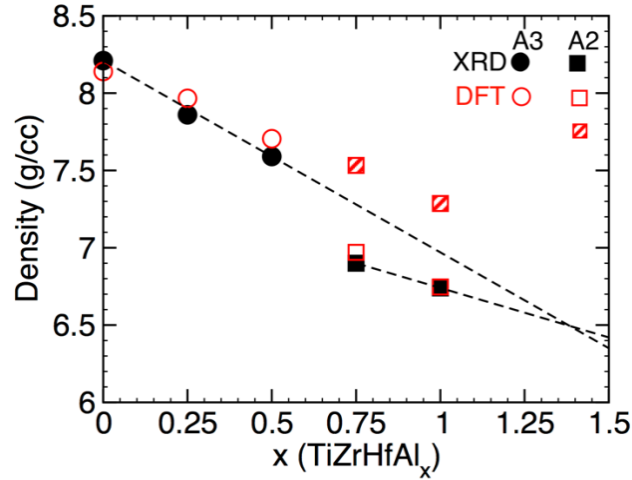

**Figure 6.** For TiZrHfAl$_x$ in A2 and A3 phases, the density (in g/cc) versus *Al* concentration (from **Table S1**): experimentally measured density calculated using X-ray (solid-black symbols) compared to the theoretical values [A2 with vacancy tetrahedron (open-squares), and also values with no vacancies (hatched-squares)]. Ignoring the vacancy-tetrahedron, the theoretical density is just linear in *Al* concentration; whereas, with the vacancy-tetrahedron included there is drop in density, as observed. Dashed lines are guides to the eye for A2 and A3 phases.

*Defect-structure stabilization*: We address *homogeneously distributed* vacancies (disordered) and *partially ordered* vacancies to mimic the new γ-brass-type structure (**Fig. 4c** and **Fig. 5**). The homogeneous $E_{form}$ (**Fig. 4**) has a crossover at ~7 *at.%Vac* with a cell (**Fig. 4c**) commensurate with the identified γ-brass variant, showing that at 7.45% vacancy state (4 vacancies per 54-atom cell on average) is unstable to this ordered structure. DFT calculated electronic density of states are shown in **Fig. 7** (denoted "disordered" and "partially-ordered"),[20] visually highlighting the relative change in band-energy. Clearly, occupied states below the Fermi energy move even lower and stabilize the case with "partially" chemical order and vacancy-cluster order, both indicated by the DFT calculated SRO.

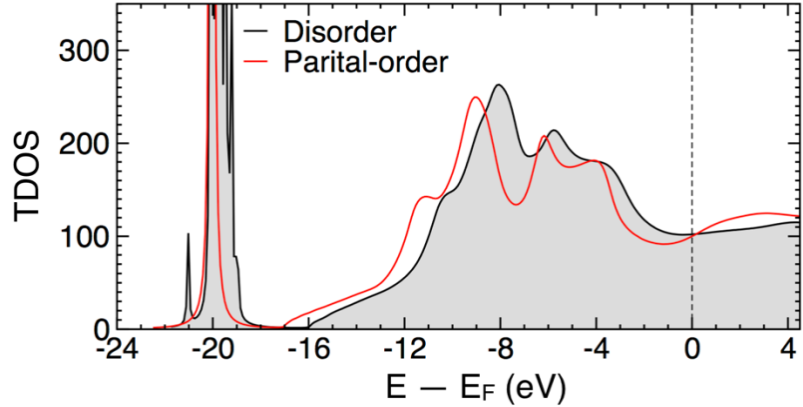

**Figure 7.** The DFT-calculated total density of states (TDOS in $states-eV^{-1}-atom^{-1}$) relative to their respective Fermi energy for bcc TiZrHfAl (54-atom) supercell with homogeneous vacancies (grey-shaded) compare to the partially-ordered γ-brass structure (**Fig. 4c**), both at experimental lattice constant (10.196 Å) to identify key electronic shifts. Chemical disorder-induced broadening is evident by smooth solid-solution DOS. For γ-brass, a lower band-energy relative to "disordered" case shows increased stability (decreased electronic energy) with vacancy ordering. States at –19.8 eV are localized Hf *f*-states.

**Conclusions**

We investigated the crystal structure of refractory-based TiZrHf-Al high-entropy alloys by an integrated combination of single-crystal diffraction and high-energy synchrotron radiation supported by DFT calculations to account for disorder and to reveal the role of vacancy-mediated order. Unexpectedly, the structure with Al additions was found to be a bcc-based variant of the well-known *γ*-brass, with 4 vacant sites (not 2 vacancies as in traditional brass). Our work highlights the importance of vacancy-mediated stability of the new variant of *γ*-brass structure, where even small amounts of vacancies suppress atomic SRO and drive vacancy ordering, providing a critical gain in energy to stabilize brass-like structure. As vacancies are inherent from processing in refractory-based systems, it is expected that similar discoveries await in other high entropy alloys, especially in systems that display the characteristics of (1) marginal high entropy alloys (those with entropy not so strong that they tend to order) for inducing weak ordering and (2) a valence-electron count that favors formation of defect-stabilized phases. Moreover, with improvements in X-ray diffraction sensitivity in the last two decades, we suspect that some older experimental data suggesting single-phase A2 phases in some systems may in fact have superstructures as found here.

## Acknowledgements

Experimental work and application of theory to this system was supported by the U.S. Department of Energy (DOE), Office of Fossil Energy, Crosscutting Research Program. The SRO linear-response theory development (PS, AVS, DDJ) was funded by the U.S. DOE Office of Science, Basic Energy Sciences, Materials Science & Engineering Division. Research was performed at Iowa State University and Ames Laboratory, which is operated by ISU for the U.S. DOE under contract DE-AC02-07CH11358. The Advanced Photon Source use was supported by U.S. DOE, Office of Science, Office of Basic Energy Sciences under Contract No. DE-AC02-06CH11357.